\documentclass{article}
\usepackage{spconf,amsmath,graphicx}
\usepackage{amsmath}
\usepackage{amsfonts}
\usepackage{multirow}
\usepackage{hyperref}
\newcommand{\norm}[1]{\left\lVert#1\right\rVert}

\title{Multi-level reversible encryption for ECG signals \\using compressive sensing}
\name{Mikko Impiö$^{\dag\star}$ \qquad Mehmet Yama\c{c}$^{\dag}$ \qquad Jenni Raitoharju$^{\star}$ \thanks{This  work  has been  supported  by  the  NSF-Business  Finland
Center for Visual and Decision Informatics (CVDI) project
AMALIA.}}
\address{$^{\dag}$ Faculty of Information Technology and Communication Sciences, Tampere University, Finland
\\$^{\star}$Programme for Environmental Information, Finnish Environment Institute, Jyväskylä, Finland}
\begin{document}
%
\maketitle
\ninept
\begin{abstract}
Privacy concerns in healthcare have gained interest recently via GDPR, with a rising need for privacy-preserving data collection methods that keep personal information hidden in otherwise usable data. Sometimes data needs to be encrypted for several authentication levels, where a semi-authorized user gains access to data stripped of personal or sensitive information, while a fully-authorized user can recover the full signal.
In this paper, we propose a compressive sensing based multi-level encryption to ECG signals to mask possible heartbeat anomalies from semi-authorized users, while preserving the beat structure for heart rate monitoring. Masking is performed both in time and frequency domains. Masking effectiveness is validated using 1D convolutional neural networks for heartbeat anomaly classification, while masked signal usefulness is validated comparing heartbeat detection accuracy between masked and recovered signals. The proposed multi-level encryption method can decrease classification accuracy of heartbeat anomalies by up to $50\%$, while maintaining a fairly high R-peak detection accuracy.
\end{abstract}
\section{INTRODUCTION}
\label{sec:intro}
Electrocardiography is an important method for detecting cardiac abnormalities, being a standard procedure performed by healthcare professionals worldwide. Increased attention on patient privacy in the recent years, especially after the EU General Data Protection Regulation (GDPR), has directed attention to privacy-preserving data collection methods in healthcare. Full electrocardiogram (ECG) data can be used as a biometric \cite{huang2019practical} and can contain sensitive information concerning cardiac arrhythmia and abnormalities. This sensitive data may be collected or handled by parties who do not necessarily need the access to the full information on health problems, but instead they only need to ensure the basic quality of the data and well-being of the patient, e.g., by monitoring the heart beat. However, when needed, fully-authorized users should be able to access the full ECG recording. At the same time, the data should be fully inaccessible to outsiders. Thus, there is a need for multi-level privacy-preserving ECG collection methods that will reveal different amounts of information to users with different authorization levels.

Privacy preservation in ECG monitoring has been widely studied in the past. For example, Huang et al. studied privacy-preserving ECG collection for usage as a biometric in \cite{huang2019practical}. Barni et al. \cite{barni2011privacy} studied ECG classification while maintaining the data privacy. Taking into account compression issues in data acquisition, Mamaghianian et al. \cite{mamaghanian_compressed_2011} first proposed compressive sensing for efficient ECG compression. Compressive sensing has proved to be a good multi-level encryption and compression method for sparse signals, having being used for images by Yamaç et al. in \cite{yamac_reversible_2019, yamac2020privacy_TIFS}. Their compressive sensing based encryption method makes it possible to simultaneously encrypt and compress the ECG signal for safe and compact transmission and provides two-level recovery for different levels of user authorization.

\begin{figure}[ht]
\centering
\begin{minipage}[b]{0.8\linewidth}
  \centering
  \centerline{\includegraphics[width=6.0cm]{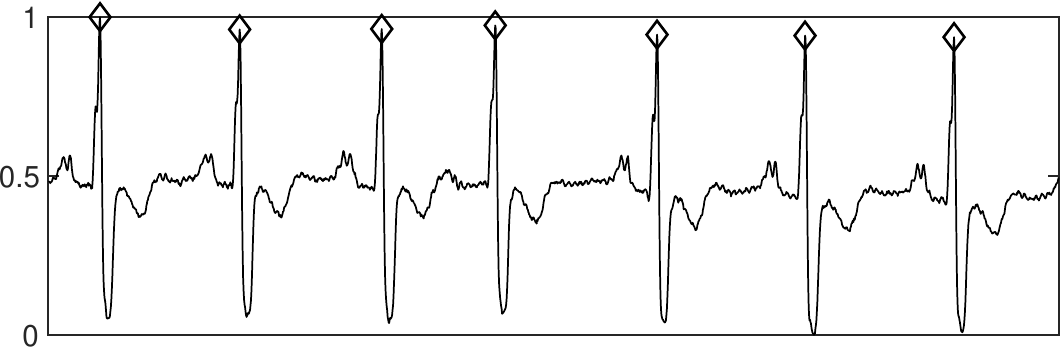}}
  \centerline{(a) Original signal, class 'S'}\medskip
\end{minipage}
\begin{minipage}[b]{0.8\linewidth}
  \centering
  \centerline{\includegraphics[width=6.0cm]{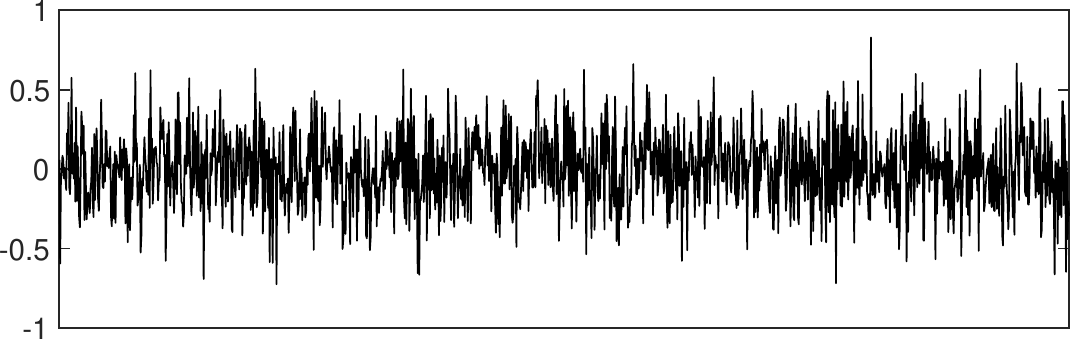}}
  \centerline{(b) Encrypted signal}\medskip
\end{minipage}
\begin{minipage}[b]{0.8\linewidth}
  \centering
  \centerline{\includegraphics[width=6.0cm]{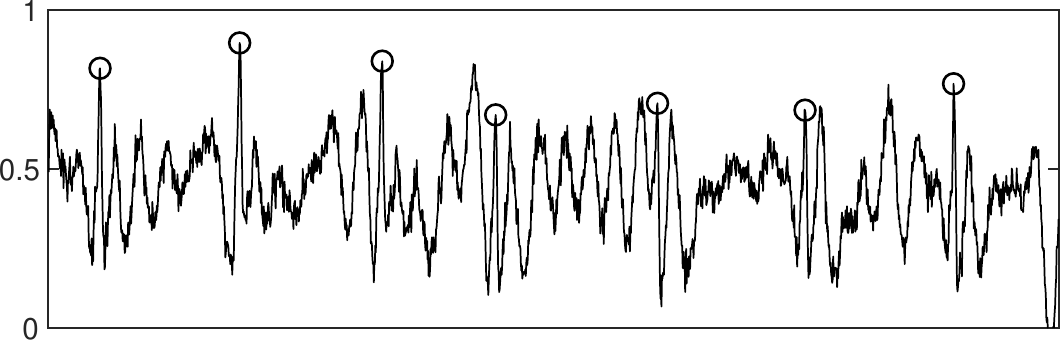}}
  \centerline{(c) Partially masked signal for User A, classified as 'V'}\medskip
\end{minipage}
\begin{minipage}[b]{0.8\linewidth}
  \centering
  \centerline{\includegraphics[width=6.0cm]{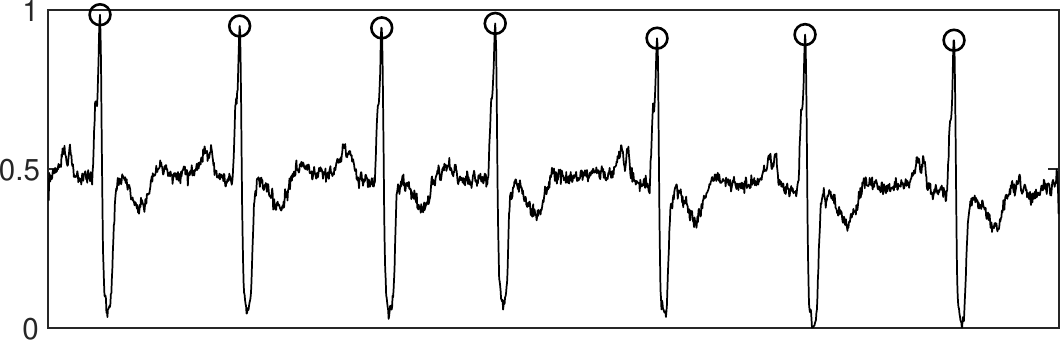}}
  \centerline{(d) Fully recovered signal for User B, classified as 'S'}\medskip
\end{minipage}
\vspace{-5pt}
\caption{Original and encrypted signals (a,b), with recovered signals for Users A and B (c,d)}
\label{fig:classification}
\end{figure}
In this paper, we propose a compressive sensing based encryption method for reversible multi-level masking of ECG signals. This is achieved by masking the original ECG signal embedding a mask on the sensitive parts of the original signal. The proposed method makes it possible to produce a multi-level masked signal during sampling, making full reconstruction possible only for authorized users. While the proposed method is similar to the reversible multi-level encryption introduced in \cite{yamac_reversible_2019, yamac2020privacy_TIFS} for images, we investigate its usefulness for the first time for 1-D signals. Furthermore, we propose
a novel approach for frequency-domain masking in order to maintain more structure in the signal for semi-authorized users. Possible use cases for the proposed method include hospital scenarios, where semi-authorized staff could monitor QRS peak information from the ECG, while keeping possible sensitive abnormalities in the ECG private. Illustrative example of the method is presented in Figure~\ref{fig:classification}, where an abnormal supraventricular (S) ECG compressed by factor $0.65$ is masked in frequency domain. This maintains its structure for accurate R-peak detection for semi-authorized User A, but hides the abnormality from a heartbeat classification algorithm and the inspector. For fully-authorized User B the signal is recovered and classified correctly. Reference beats are marked with diamonds, beat detection results on compressed signals with circles.

\section{BACKGROUND}
\label{sec:methods}
\subsection{Compressive sensing}

Compressive sensing \cite{candes2006compressive} has been introduced as an alternative to the traditional Nyquist-Shannon sampling paradigm, as sparse signals can be sampled using a lower sampling rate and recovered without significant loss in quality. In compressive sensing, the objective is to represent a signal $\mathbf{s}\in \mathbb{R}^N$ uniformly sampled using the traditional data acquisition scheme can be with $m << N$ measurements as $\mathbf{y} = \mathbf{A}\mathbf{s}$,
where $\mathbf{A}\in \mathbb{R}^{mxN}$ is a compressive sensing measurement matrix, and $m$ is specified by a measurement rate (MR) of the original signal. Most natural signals are sparse in some proper domain $\mathbf{\Phi}$, i.e., they can be accurately given as $\mathbf{s} = \mathbf{\Phi}\mathbf{x}$, where $\mathbf{\Phi}$ is $\mathbb{R}^{NxN}$ sparsifying basis and $\mathbf{x}$ contains less than $k<N$ non-zero coefficients. Under this sparsity constraint, the underdetermined system of equations $\mathbf{y} = \mathbf{A}\mathbf{s}$ can we rewritten as
\begin{equation}
\label{eq:sensingmatrix}
    \mathbf{y} = \mathbf{A}\mathbf{\Phi}\mathbf{x} = \mathbf{Hx} \quad \text{s.t.} \quad \norm{\mathbf{x}}_0 \leq k.
\end{equation}
The goal is to find the sparsest solution where the amount of lost information is less than $\epsilon$: 
\begin{equation}
\label{eq:basispursuit0}
    \hat{\mathbf{x}} = \underset{\mathbf{x}}{\mathrm{argmin}}\norm{\mathbf{x}}_0 \quad \text{s.t.} \quad \norm{\bf{Hx} - \bf{y}}_2 \leq \epsilon, 
\end{equation} 
This non-convex optimization is typically relaxed to the $\ell_1$-norm based optimization problem
\begin{equation}
\label{eq:basispursuit}
    \hat{\mathbf{x}} = \underset{\mathbf{x}}{\mathrm{argmin}}\norm{\mathbf{x}}_1 \quad \text{s.t.} \quad \norm{\bf{Hx} - \bf{y}}_2 \leq \epsilon, 
\end{equation}
The optimization routine is a well-known and widely studied problem for recovery of compressively sensed signals.

\subsection{Multi-level encryption}

Prior research on using compressive sensing as an encryption system has shown that without knowing the measurement matrix $A$, it is hard to recover information in a feasible time \cite{rachlin2008secrecy}, making it a reliable encryption scheme. It also has been shown in \cite{bianchi2015analysis} that if a random i.i.d. Gaussian matrix is used as the measurement matrix $A$, no information can be recovered from the original signal apart from its energy. Yamaç et. al. \cite{yamac_reversible_2019, yamac2020privacy_TIFS} proposed a multi-level encryption method, 
where the sensitive parts of the signal are masked with an additive perturbation mask $\mathbf{M}$
\begin{equation}
\label{eq:perturbation}
\bf{y_d} = (\bf{A}+\bf{M})\bf{s}.
\end{equation}
Outsiders not knowing the measurement matrix $\mathbf{A}$ cannot recover the signal at all, while semi-authorized users (User A) can recover the degraded signal in \eqref{eq:perturbation}. For fully-authorized users (User B), additional information about the mask is needed. The perturbation matrix $\mathbf{M}$ can be encoded into a vector $\mathbf{w}$ of length $T<m$ and this information can be watermarked directly into the signal using an embedding matrix $\mathbf{B}\in\mathbb{R}^{m\times T}$ which is known only to fully-authorized users:
\begin{equation}
\label{eq:embedded}
\bf{y_w} = (\bf{A}+\bf{M})\bf{s} + \bf{Bw} = \bf{Hx} + \bf{Bw} + \bf{n},
\end{equation}
where $ \mathbf{H}\mathbf{x} = \mathbf{A} \mathbf{\Phi} \mathbf{x} = \mathbf{A} \mathbf{s}$ and $\mathbf{x} \in \mathbb{R}^N$ is the sparse representation of $\mathbf{s}$ in $\mathbf{\Phi}$, and the masked part can be expressed as noise term, i.e., $ \mathbf{n} = \mathbf{M} \mathbf{s}$. The watermark is multiplied with an embedding power $a$, where higher embedding power amounts to higher energy additive noise in the signal, but makes the full recovery of the watermark more reliable.

User A can recover the degraded signal following the normal compressive sensing reconstruction approach and ignoring the noise. User B can recover the signal by removing the embedded information using the left annihilator matrix of $\bf{B}$,  $\bf{F}\in \mathbb{R}^{P\times m}$ ,  where $P = m-T$, as $\hat{\mathbf{y}} = \mathbf{Fy_w}$ and solving
\begin{equation}
\label{eq:x_pre_estimation}
    \tilde{\mathbf{x}} = \underset{\bf{x}}{\mathrm{argmin}}\norm{\mathbf{x}}_1 \quad \text{s.t.} \quad \norm{\bf{FHx} - \bf{\tilde{y}}}_2 \leq \epsilon, 
\end{equation}
where $\tilde{y} = \bf{F(Hx + Bw + n)} = \bf{FHx+Fn}$, $\bf{Fn}$ being additive noise on the signal.

Using this pre-estimate, the watermark can be estimated using the pseudo-inverse (least-squares) of $\bf{B}$:
\begin{equation}
\bf{w''} = (\bf{B}^T\bf{B})^{-1}\bf{B}^T(\bf{y_w-H\tilde{x}}),
\end{equation}which can be furtherly thresholded to produce an improved estimate $\bf{\hat{w}}$. Finally, the watermark estimate is used to recover an estimate of the additive perturbation mask $\bf{\hat{M}}$. $\bf{\hat{M}}$ and $\bf{\hat{w}}$ can be used to recover the original signal:
\begin{equation}
\label{eq:userB}
\bf{\hat{x}} =  \underset{\bf{x}}{\mathrm{argmin}}\norm{x}_1  \text{ s.t. } \norm{(\bf{y_w-B\hat{w}}) - (\bf{A+\hat{M})\Phi x} }_2 \leq \epsilon.
\end{equation}
The original signal is recovered simply by applying the transform $\bf{\hat{y}} = \bf{\Phi\hat{x}}$. The full algorithms for encryption and recovery under this multi-level encryption scheme can be found \cite{yamac_reversible_2019, yamac2020privacy_TIFS}, while \cite{yamac2020privacy_TIFS} provides also theoretical analysis of the cryptographic properties. 

\section{PROPOSED METHOD}

In this paper, we propose the above-described multi-level encryption approach of for ECG signals. Previously, the method has been used only for images. We first formulate a time-domain masking scheme, which is similar to the masking in \cite{yamac_reversible_2019}. Furthermore, we propose a novel approach for frequency-domain masking. 

\subsection{Time-domain masking}

Masking the signal $\mathbf{s}$ for semi-authorized User A is performed using a perturbation matrix $\bf{M}$. The perturbation matrix is formed using a mask vector $\bf{c}$ containing the information of the signal indices to be masked. We select the signal indices by detecting R-peaks using the Pan-Tompkins algorithm \cite{pan1985real}, and masking a fixed distance from each peak index.



\begin{figure}[t]
    \centering
    \includegraphics[width=1.0\linewidth]{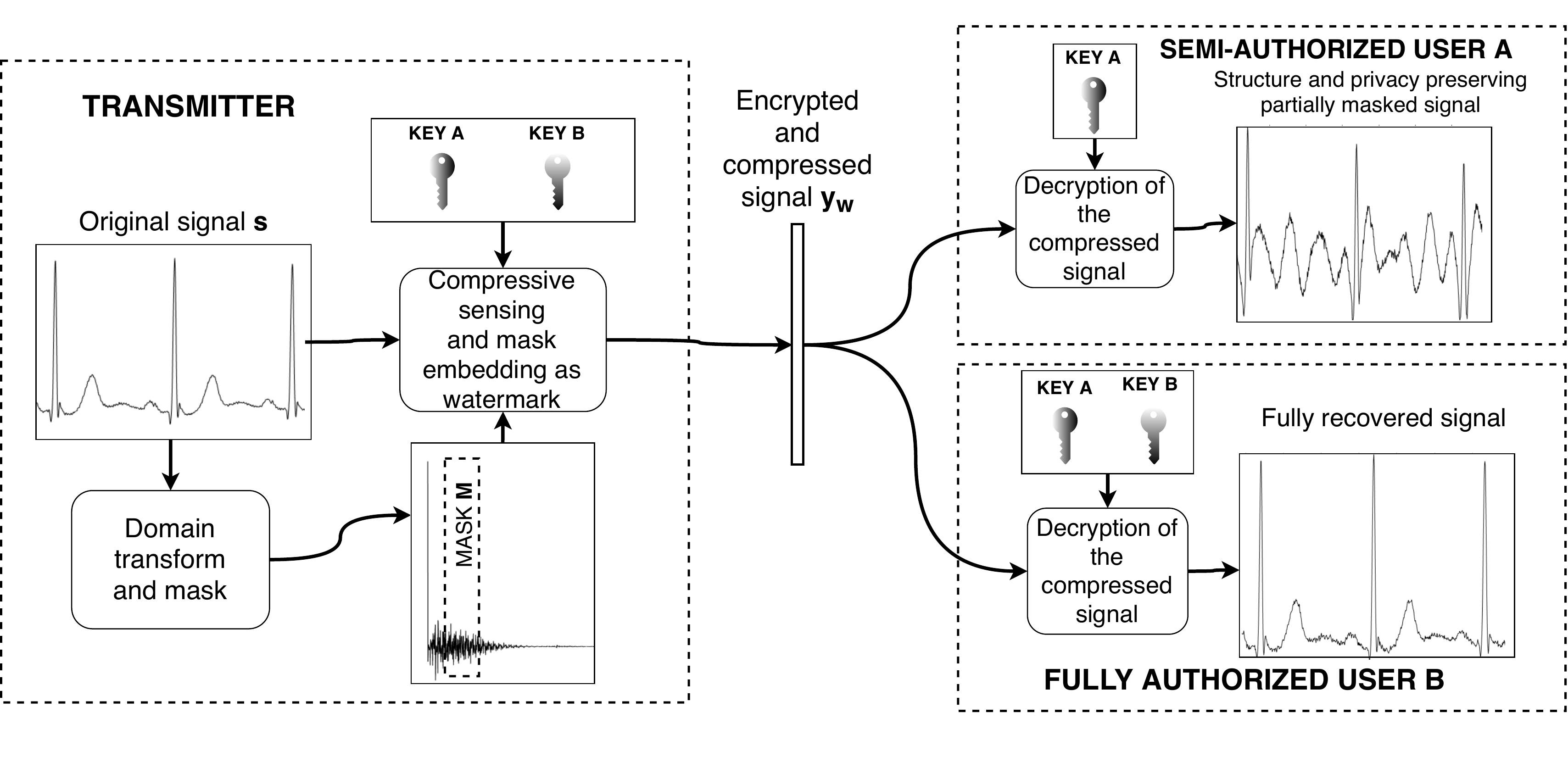}
    \caption{Diagram of the multi-level encryption-decryption process}
    \label{fig:flowchart}
    \vspace{-5pt}
\end{figure}

After defining the indices to be masked, we create a vector $\mathbf{p}$ of length $m$ of values $\pm 1$ following a Bernoulli distribution with some probability $p$. The additive perturbation matrix $\bf{M}$ is then formed as distribution:
\begin{equation}
\label{eq:mask_forming}
m_{i,j} = 
\begin{cases}
0, & \text{if}  \ c_j = 1 \text{ and } p_i = -1 \\
-2A_{i,j}, & \text{if}  \ c_j = 1 \text{ and } p_i = 1\\
0, & \text{if} \ c_j = 0 
\end{cases}
\end{equation}
This information can be encoded to a watermark $\bf{w}$ of length $T<m$ as
\begin{equation}
\label{eq:w}
w_{i} = 
\begin{cases}
a , & \text{if} \ i \leq l \text{ and } p_i = 1 \\
-a, & \text{if} \ i \leq l \text{ and } p_i = -1 \\
0, & \text{if} \ c_j = 0, 
\end{cases}
\end{equation}
where $l$ is amount of indices to be masked and $a$ is the embedding power.
This watermark vector can optionally be further encoded to a different ternary representation for compression and robustness of signal transmission, assuming the receiver can decode the embedded watermark. The full approach for multi-level ECG signal encryption, transmission and decoding procedure is described in Figure \ref{fig:flowchart}.

\subsection{Frequency-domain masking}



In frequency-domain, a certain frequency band is masked. The frequency band range can be known to the receiver, so only the mask vector values need to be embedded.
If the original measurement matrix is $\bf{A}$, we mark the frequency-domain measurement matrix as $\bf{A}_f = \bf{A}\Omega^T$. Due to the masking is done in transformation domain with basis $\mathbf{\Omega}$, the measurement matrix has to contain this information. The sensing matrix is now $\bf{H} = \bf{A}_f\bf{\Phi} = \bf{A}\Omega^T\Phi$. The frequency-domain counterpart for \eqref{eq:embedded} is now
\begin{equation}
\label{eq:freq_embedded}
\bf{y_w} = (\bf{A}+\bf{M})\Omega^T\bf{s} + \bf{Bw} = \bf{Hx} + \bf{Bw} + \bf{n},
\end{equation}
where $\mathbf{n} = \mathbf{M\Omega^Ts}$. The rest of the recovery process is similar as in time-domain. Note that during sensing the matrix $\bf{A }$ is similar to the time-domain counterpart, while Users A and B need to have $\bf{A}_f$ for signal recovery. The frequency domain counterpart for \eqref{eq:userB} is
\begin{equation}
\label{eq:userB_freq}
\bf{\hat{x}_B} =  \underset{\bf{x}}{\mathrm{argmin}}\norm{x}_1 \text{ s.t. } \norm{(\bf{y_w-B\hat{w}}) - (\bf{A_f\Omega+\hat{M})\Phi x} }_2 \leq \epsilon.
\end{equation}

\section{EXPERIMENTS}
\label{sec:experiments}
The proposed ECG masking method was evaluated using the MIT-BIH arrhythmia database \cite{goldberger2000physiobank}. Each of the 48 patient records in the database was truncated to uniform length of $2^{18}$ samples, which were then split to 128 independent 2048 sample long signals. Each signal was labeled according to the AAMI EC57:1998 standard \cite{kandala2019towards}, where heartbeats are classified to normal (N), supraventricular (S), ventricular (V),  fusion (F), and unknown (Q) beats. Label N was assigned if all beats on the sample were normal and label S,V,F, or Q if any of these beat types were present. If several different abnormal beat types were present in the signal, the most common one was selected as the label. Table \ref{tab:AAMI_labels} presents the total amounts of samples in each class and their corresponding labels in the MIT-BIH-dataset. From this dataset, 60 signals from each N,S,V, and F types were randomly sampled for the final dataset to be masked. Codes are publicly available at \url{https://github.com/mikkoim/ECG-CS}.

\begin{table}[tb]
    \centering
    \begin{tabular}{c|c|c}
        \textbf{AAMI label} &  \textbf{MIT-BIH labels} & \textbf{N} \\\hline\hline
         N         &   1,2,3,11,34   & 4451 \\
         S         &   4,7,8,9       & 333 \\
         V         &      5,10       & 1291 \\
         F         &       6         & 62 \\
         Q         &      13         & 7 \\\hline
         Total     &                 & 6144 \\
    \end{tabular}
    \caption{Split and relabeled MIT-BIH dataset}
    \label{tab:AAMI_labels}
\end{table}

The watermark embedding power was chosen to produce the minimum amount of noise in User A recovered signal, while maintaining error-free recovery of the mask from the watermark. Embedding power was chosen empirically as the minimum value that recovers $\sim90\%$ of the masks in the dataset. Embedding powers, measurement rates and mask lengths are displayed in Table \ref{tab:em_ratio}. Time domain masking is the least efficient one due to varying locations and amounts of peaks that need to be embedded to the signal.

\begin{table}[t]
    \footnotesize
    \centering
    \begin{tabular}{l|l|l|r}
         \textbf{Mask type }         & \textbf{MR }& \textbf{EP}   & {$T$} \\\hline\hline
         Freq.         & 0.3                   &           0.5     & 110 \\
         Freq.         & 0.5                   &           0.2     & 110 \\
         Freq.          & 0.65                  &           0.1     & 110 \\\hline
         Fixed freq.   & 0.3                   &           0.1     & 80 \\
         Fixed freq.   & 0.5                   &           0.2     & 80 \\
         Fixed freq.   & 0.65                  &           0.05    & 80 \\\hline
         Time        & 0.3                   &           4.5     & 500\\
         Time        & 0.5                   &           1.5     & 500 \\
         Time        & 0.65                  &           1.0     & 500 \\
         
    \end{tabular}
    \caption{Measurement rates (MR), embedding powers (EP), and minimum watermark lengths ($T$)}
    \label{tab:em_ratio}
\end{table}


\begin{table*}[ht]
\scriptsize
\setlength{\tabcolsep}{5pt}
\centering
\caption{Mean and standard deviation of 5-fold cross-validated heartbeat anomality classification results for three different masking schemes}
\label{tab:classification}

\begin{tabular}{ll|lll|lll|lll}
                                  &             & \multicolumn{3}{l|}{\textbf{Accuracy}}                                                    		&\multicolumn{3}{l|}{\textbf{Precision}}                                    									& \multicolumn{3}{l}{\textbf{Recall}}                                                      \\
\textbf{Type}                     & \textbf{MR} & \multicolumn{1}{l}{User A} & \multicolumn{1}{l}{User B} 			& \multicolumn{1}{l|}{Difference} & \multicolumn{1}{l}{User A} 			& \multicolumn{1}{l}{User B} 		& \multicolumn{1}{l|}{Difference} & \multicolumn{1}{l}{User A} 			& \multicolumn{1}{l}{User B} 		& \multicolumn{1}{l}{Difference} \\ \hline\hline
\multirow{3}{*}{Freq.}       & 0.3         & 0.2625    $\pm$0.03        & 0.5542  $\pm$0.08                  	& \textbf{0.2917}                 & 0.1390    $\pm$0.09                & 0.6811 $\pm$0.06                   & \textbf{0.5421}                 & 0.2625  $\pm$0.03                  & 0.5542  $\pm$0.08                  & \textbf{0.2917}                \\
                                  & 0.5         & 0.3458    $\pm$0.03        & 0.7792  $\pm$0.06                  	& \textbf{0.4333}                 & 0.3459    $\pm$0.09                & 0.8130 $\pm$0.05                   & \textbf{0.4670}                 & 0.3458  $\pm$0.03                  & 0.7792  $\pm$0.06                  & \textbf{0.4333}                \\
                                  & 0.65        & 0.5292    $\pm$0.05        & 0.7875  $\pm$0.08                 	& \textbf{0.2583}                 & 0.6508    $\pm$0.12                & 0.8166 $\pm$0.07                   & \textbf{0.1657}                 & 0.5292  $\pm$0.05                  & 0.7875  $\pm$0.08                  & \textbf{0.2583}                \\\hline
\multirow{3}{*}{Fixed freq.} & 0.3         & 0.2583    $\pm$0.03        & 0.5333  $\pm$0.09                 	& \textbf{0.2750}                 & 0.1104    $\pm$0.06                & 0.6713 $\pm$0.05                   & \textbf{0.5609}                 & 0.2583  $\pm$0.03                  & 0.5333  $\pm$0.09                  & \textbf{0.2750}                \\
                                  & 0.5         & 0.4417    $\pm$0.05        & 0.7750  $\pm$0.08                 	& \textbf{0.3333}                 & 0.5497    $\pm$0.23                & 0.8093 $\pm$0.05                   & \textbf{0.2596}                 & 0.4417  $\pm$0.05                  & 0.7750  $\pm$0.07                  & \textbf{0.3333}                \\
                                  & 0.65        & 0.5000    $\pm$0.09        & 0.7750  $\pm$0.08                 	& \textbf{0.2750}                 & 0.5882    $\pm$0.10                & 0.8076 $\pm$0.05                   & \textbf{0.2194}                 & 0.5000  $\pm$0.09                  & 0.7750  $\pm$0.07                  & \textbf{0.2750}                \\\hline
\multirow{3}{*}{Time}        & 0.3         & 0.2375    $\pm$0.02        & 0.3542  $\pm$0.04                  	& \textbf{0.1167}                 & 0.0959    $\pm$0.03                & 0.4809 $\pm$0.10                   & \textbf{0.3850}                 & 0.2375  $\pm$0.02                  & 0.3542  $\pm$0.04                  & \textbf{0.1167}                \\
                                  & 0.5         & 0.2583    $\pm$0.02        & 0.7542  $\pm$0.04                  	& \textbf{0.4958}                 & 0.1208    $\pm$0.09                & 0.7940 $\pm$0.04                   & \textbf{0.6732}                 & 0.2583  $\pm$0.02                  & 0.7542  $\pm$0.04                  & \textbf{0.4958}                \\
                                  & 0.65        & 0.3833    $\pm$0.05        & 0.7458  $\pm$0.06                  	& \textbf{0.3625}                 & 0.4510    $\pm$0.12                & 0.7871 $\pm$0.06                   & \textbf{0.3361}                 & 0.3833  $\pm$0.05                  & 0.7458  $\pm$0.06                  & \textbf{0.3625}                \\ \hline
\textbf{Mean}                     &             & \textbf{0.3574}            & \textbf{0.6731}            			& \textbf{0.3157}                 & \textbf{0.3391}            			& \textbf{0.7401}            		& \textbf{0.4010}                 & \textbf{0.3574}            			& \textbf{0.6731}            		& \textbf{0.3157}                \\ \cline{1-11}
\textbf{Reference}                &             & &\textbf{0.8000}                     			&            & &\textbf{0.8242}                     				& 		          					& &\textbf{0.8000}                    				&  
\vspace{-10pt}
\end{tabular}
\end{table*}

\begin{table*}[!ht]
\scriptsize
\centering
\caption{R-peak detection and mean PSNR results}
\label{tab:peak}
\begin{tabular}{ll|l|ll|ll|ll}
           &  &  & \multicolumn{2}{l|}{\textbf{Precision}}                           & \multicolumn{2}{l|}{\textbf{Recall}}                              & \multicolumn{2}{l}{\textbf{PSNR}}                                \\
\textbf{Type}       & \textbf{MR}   & \textbf{Full recovery rate}        & User A & User B & User A & User B & User A & User B \\\hline\hline
\multirow{3}{*}{Freq.}     & 0.3                    & 0.9352                      & 0.0739                     & 0.2075                     & 0.2776                     & 0.7404                     & 13.5590                    & 23.9859                    \\
                         & 0.5                    & 0.9636                      & 0.1239                     & 0.9265                     & 0.4927                     & 0.9837                     & 16.4144                    & 33.1298                    \\
                         & 0.65                   & 0.9717                      & 0.4200                     & 0.9807                     & 0.8308                     & 0.9912                     & 17.3404                    & 39.2938                    \\\hline
\multirow{3}{*}{Fixed freq.} & 0.3                    & 0.9879                      & 0.0748                     & 0.1817                     & 0.3042                     & 0.7180                     & 13.5885                    & 23.9017                    \\
                                 & 0.5                    & 0.8826                      & 0.2873                     & 0.9437                     & 0.7458                     & 0.9801                     & 17.0971                    & 33.7773                    \\
                                 & 0.65                   & 0.8907                      & 0.8019                     & 0.9842                     & 0.9163                     & 0.9920                     & 17.3138                    & 39.6812                    \\\hline
\multirow{3}{*}{Time}  & 0.3                    & 0.7247                      & 0.0815                     & 0.0908                     & 0.1163                     & 0.1296                     & -2.2180                    & 14.2293                    \\
                             & 0.5                    & 0.9069                      & 0.2712                     & 0.3422                     & 0.3696                     & 0.5507                     & 5.3263                     & 25.1628                    \\
                              & 0.65                   & 0.8947                      & 0.9355                     & 0.8888                     & 0.7275                     & 0.9660                     & 11.2433                    & 35.4876                   
\end{tabular}
\end{table*}

Three different masking methods were evaluated: (1) time-domain masking, (2) frequency-domain masking for a single frequency band interval and (3) frequency-domain masking for a predefined interval. The fixed mask allows for lower embedding powers, as can be seen from Table \ref{tab:em_ratio}, with the trade-off of slightly degraded security. The peak detection mask was set to 15 samples around the detected R-peak for every heartbeat. Both frequency domain masks were set empirically to the frequency range $[20,90]$ that minimized classification accuracy and maximized R-peak detection accuracy. Discrete cosine transform was used for both $\mathbf{\Phi}$ and $\mathbf{\Omega}$ transforms. Wavelet transform was used as the basis $\mathbf{\Phi}$ for time-domain signals. Basis-pursuit optimization for all mask types was performed using the $\ell_1$-magic -optimization routines \cite{candes2005l1}. For the measurement matrix $\bf{A}$ and embedding matrix $\bf{B}$, random Gaussian matrices were used due to them being close to optimal for satisfying the restricted isometric property for compressive sensing \cite{candes2006near}.

\subsection{Heartbeat classification}

The masking efficiency in both time and frequency-domains was evaluated using a 1D-CNN based on the ResNet \cite{he2016deep} architecture. 1D-CNN-based heartbeat classification algorithms have been used for example in \cite{yildirim2018arrhythmia, li2017classification, kiranyaz2015real}, while the ResNet variant has been proposed by \cite{andreotti2017comparing, xiong2018ecg}. The used 1D ResNet implementation is based on the one used by Hong et al. in \cite{hong2017encase}. A 50-layer 1D ResNet was used as the base model for the ECG classification. This model was first pre-trained for 10 epochs using 5897 samples not present in the final dataset.

The final dataset of $4\cdot 60=240$ samples was split to training (80\%) and testing (20\%) datasets for 5-fold cross-validation. For each fold, a separate model was trained for 50 epochs by continuing the training of the base model using the balanced final dataset. Each model was validated using equivalent samples of the original signal and its User A and B counterparts. Mean accuracy, precision and recall were calculated over the 5 folds and are presented in Table \ref{tab:classification} along with the standard deviation of each metric.

It can be seen from Table \ref{tab:classification} that higher measurement rate (MR) results in higher classification accuracies for both signals. However, for each masking method and measurement rate, the decrease in classification accuracy is significant, resulting in increased privacy. User A classification accuracy is close to or even below the random case for lower measurement rates, while User B classification accuracy is close to the reference classifier for higher measurement rates. The best results for multi-level ECG masking were achieved using the stationary frequency-domain mask, with measurement rate of 0.65. 

In general, User A signals are less accurately classified, arguably both due to the masking, but also due to the masked signals not being from a similar distribution as the 'normal' training data. Some initial experiments were conducted by training the classifier with User A masked data, achieving a higher accuracy due to the train and test samples being from the same distribution. This could pose some privacy concerns if an semi-authorized user has access to a large dataset of masked signals with known true labels. However, in a real environment this is an unlikely scenario.

\subsection{QRS peak detection}

Further analysis on the usability of partially masked (User A) and fully recovered (User B) ECG signals was performed by calculating the beat detection accuracy and peak-signal-to-noise-ratios (PSNR) for different masking methods. Widely used Pan-Tompkins QRS-complex R-peak detection algorithm \cite{pan1985real} was used on the original signal to produce reference points for beat detection. Same algorithm was used on User A and B signals, where a detection was considered correct if a single detected peak was marked within 10 samples from a reference peak. Peak detection precision and recall scores are presented in Table \ref{tab:peak}.

Table \ref{tab:peak} also presents the rate of flawless mask recovery and the mean PSNR values for both user types for each masking method. Full recovery rate states the percentage of signals without any errors in mask recovery. Possible errors in mask recovery result in degradation of User B signal, when the recovered signal does not match the original. Setting the embedding power higher improves the full mask recovery rate, with the trade-off of additional noise in the User~A signal. The embedding power in these experiments was set so that $\sim 90\%$ full recovery rate is achieved. Due to the low sampling rate of measurement rate 0.3 for time-domain masking, this could not be achieved, resulting in unusable PSNRs for both User A and B for low MRs.

\section{CONCLUSIONS}
\label{sec:conclusion}

In this paper, we proposed a multi-level compressive sensing-based encryption for ECG signals for different levels of user authorization. The conducted ECG classification experiments indicated that the proposed scheme can hide the sensitive information on heart problems from semi-authorized users, while keeping it possible to monitor heart rate. At the same, the proposed approach allows fully-authorized users to recover the original signal information.

\newpage

\bibliographystyle{IEEEbib}
\bibliography{strings,refs}

\end{document}